 \documentclass[authoryear,preprint,review,12pt]{elsarticle}

\usepackage{color}

\usepackage{graphicx}
\usepackage{lineno}  
\usepackage{amssymb}
\usepackage{float}
\usepackage{arydshln}
\usepackage{url}

\usepackage{multirow}
\usepackage{threeparttable}
\usepackage{bm}
\usepackage{amsmath}
\usepackage{algorithm}
\usepackage{algorithmic}

\usepackage[x11names]{xcolor}
\usepackage[
  colorlinks,
]{hyperref}

\AtBeginDocument{\hypersetup{citecolor=DodgerBlue4}}

\journal{elsevier}

\begin{document}

\begin{frontmatter}
\title{Modeling shared micromobility as a label propagation process for detecting the overlapping communities}

\author[a]{Peng Luo}
\ead{pengluo@mit.edu}

\author[b]{Chengyu Song}
\ead{ge95mol@mytum.de}

\author[c]{Hao Li}
\ead{hao_bgd.li@tum.de}

\author[d]{Di Zhu}
\ead{dizhu@umn.edu}

\author[a]{Songhua Hu}
\ead{hsonghua@mit.edu}

\author[a]{Fábio Duarte}
\cortext[cor1]{Corresponding author: Fábio Duarte, fduarte@mit.edu}

\address[a]{Senseable City Lab, Massachusetts Institute of Technology, Cambridge, USA}

\address[b]{Chair of Cartography and Visual Analytics, Technical University of Munich, Munich, Germany}

\address[c]{Professorship of Big Geospatial Data Management, Technical University of Munich, Munich, Germany}

\address[d]{Department of Geography, Environment and Society, University of Minnesota, Twin Cities, Minneapolis, USA}

\begin{abstract}

Shared micro-mobility such as electric scooters (e-scooters) has gained significant popularity in many cities. While many studies have analyzed the spatiotemporal patterns of shared micro-mobility using individual-level trip data, the spatial structure of e-scooter mobility networks and their socio-economic implications remain underexplored. Examining these mobility networks through the lens of network science—such as analyzing their community structures—can provide valuable insights for urban policy and planning. For example, allocating e-scooters at the overlapping locations of two communities may improve the operational efficiency of e-scooter distribution. However, existing methods for detecting community structures in mobility networks often overlook potential overlaps between communities. In this study, we conceptualize shared micro-mobility in urban spaces as a process of information exchange, where locations are connected through e-scooters, facilitating the interaction and propagation of community affiliations. As a result, similar locations are assigned the same label. Based on this concept, we developed a Geospatial Interaction Propagation model (GIP) by designing a Speaker-Listener Label Propagation Algorithm (SLPA) that accounts for geographic distance decay, incorporating anomaly detection to ensure the derived community structures reflect meaningful spatial patterns.We applied this model to detect overlapping communities within the e-scooter system in Washington, D.C. The results demonstrate that our algorithm outperforms existing model of overlapping community detection in both efficiency and modularity. Additionally, we discovered significant social segregation within the overlapping communities: areas belong to multiple communities tend to be wealthier with shorter commute times. Our results provide a potential explanation for the community structure in human mobility networks and may offer insights for urban planning and policymaking aimed at creating a more equitable and accessible mobility system.

\end{abstract}

\end{frontmatter}


\section{Introduction}

Over the past decade, the emergence of shared micro-mobility options, such as (electric) bicycles and scooters, has influenced human travel behaviors in urban spaces \citep{rose2012bikes,li2022understanding}. Numerous studies using individual-level trip data from shared micro-mobility systems have focused on various aspects, including spatiotemporal analysis \citep{chen2024complementary, hu2022examining}, travel behavior characteristics \citep{guo2021understanding, jing2021joint}, transportation efficiency improvement \citep{moreau2020dockless,wanganoo2022intelligent}, environmental benefits \citep{badia2023shared}, and transportation equality \citep{laa2020survey,nikiforiadis2021analysis,hu2021examining}. However, individual trip-based analyses often fall short in uncovering the clustering patterns and spatial structures inherent in micro-mobility systems. These clustering analyses are crucial as they expose hidden socioeconomic dynamics, such as the ways different social groups interact with urban spaces and how these interactions mirror transportation inequalities and social segregation \citep{luo2022sensing}. Moreover, the can guide the deployment \citep{zhang2022allocation}, management, and optimization of emerging transportation modes like e-scooters \citep{song2021spatiotemporal}. Despite recent advancements in shared micro-mobility, a gap remains in understanding the spatial patterns of these systems and their associations with socio-spatial factors \citep{hosseinzadeh2021spatial}.

Community detection, a clustering analysis method grounded in network science, is an effective approach for uncovering spatial patterns in human mobility, including shared micro-mobility. Community detection involves partitioning the nodes of a network into clusters, where nodes within the same cluster exhibit dense interconnections, while connections between clusters are relatively sparse \citep{gulbahce2008art}. Mobility networks can be conceptualized as spatial graphs, where origins and destinations are represented as nodes, and trip flows between them are represented as edges \citep{zhu2020understanding}.By identifying locations with frequent and strong mobility interactions as communities \citep{hossmann2011complex,thuillier2017clustering,rinzivillo2012discovering}, this approach provides insights into mobility-driven spatial structures. Unlike predefined administrative boundaries, these communities emerge organically from actual mobility behaviors and spatial connections, offering a more nuanced understanding of urban dynamics. 

In current research on community detection within human mobility network, the primary approach is based on modularity optimization algorithms \citep{lin2020revealing}. Modularity optimization aims to identify densely connected groups of nodes, or communities, by maximizing the difference between internal connectivity within the community and external connectivity to other communities \citep{newman2006modularity}. Methods based on these algorithms have been applied to new transportation systems, such as bike-sharing or e-scooter sharing, with approaches including the Louvain algorithm \citep{shi2019finding} and the LEIDEN algorithm \citep{chen2022delineating}. Moreover, one study \citep{kara2024comprehensive} conducted a hierarchical and dynamic community detection study on the bike-sharing system in Munich, utilizing the Infomap method.

Although the aforementioned methods have explored community detection in shared mobility networks, they primarily focus on identifying disjoint communities, where each location is assigned to a single, specific community. However, communities within shared mobility networks can often overlap \citep{wang2024overlapping}. Given the complexity of mobility flows in shared mobility systems, certain locations may simultaneously belong to multiple communities. For example, e-scooter users from different residential areas might travel to the same supermarket or school. In this case, the supermarket serves as an overlapping location for two distinct e-scooter communities. Identifying and analyzing such overlapping communities in urban environments is critical for optimizing transportation systems, urban planning, and resource allocation. For instance, allocating more e-scooters to overlapping locations can increase utilization rates, serve a larger number of users, and reduce the distance required for e-scooter redistribution. This, in turn, helps lower emissions and contributes to a more sustainable urban transportation system.
Previous research has already revealed overlapping community structures in general human mobility networks \citep{luo2022sensing,luo2024uncover}, highlighting their strong connections to socioeconomic factors. 

In this study, we developed an overlapping community detection algorithm and applied it to e-scooter systems. We conceptualized trip flow within e-scooter systems as an interaction propagation process, where trips between different locations represent spatial connections. Each location is assigned a label representing its community affiliation (For example, if location \textit{i} belongs to community \textit{A}, then \textit{A} serves as the label of \textit{i}), and during the propagation process, label information is communicated, exchanged, or updated. Ultimately, locations with strong connections will converge to share the same label, indicating membership in a common community. Based on this concept, we proposed a Geospatial Interaction Propagation (GIP) model for detecting overlapping communities in human mobility networks. By combining the Speaker-Listener Label Propagation Algorithm (SLPA) \citep{xie2011slpa} with a one-class support vector machine (SVM) \citep{manevitz2001one,li2003improving}, our approach accounts for both distance decay and mobility connections between locations, effectively filtering out communities with minimal spatial interaction and improving detection accuracy.

We applied the GIP model to detect overlapping communities within Washington, D.C.'s e-scooter network. First, we performed a spatial join between the e-scooter origin-destination (OD) data and Census Block Groups (CBGs). Second, we conceptualized the e-scooter network as a dual graph, where CBGs serve as nodes, connections between them represent edges, and edge weights indicate the intensity of human mobility between origin and destination CBGs. Finally, we used the SLPA algorithm enhanced by one-class SVM to group CBGs into distinct communities. After identifying the overlapping communities, we conducted a systematic analysis to understand their characteristics including transportation services, land use distribution (measured by the co-occurrence of Points of Interest (POIs)), and sociodemographics (measured by income, employment status, and racial makeup).

The remainder of this paper is structured as follows: Section 2 introduces the study area and datasets. Section 3 describes the GIP model. Section 4 presents the case study using the GIP model to detect overlapping communities. We discuss the findings in Section 5 and conclude the study in Section 6.

\section{Data and study area}

\subsection{Study area}

Washington, D.C. was selected as our study area due to its pioneering and well-developed shared micro-mobility systems in the U.S. \citep{ddot2018} It is also one of the busiest cities in the U.S. in terms of human mobility, urban vitality, and commuting traffic \citep{acs2006}. Managed under the District Department of Transportation (DDOT), the shared micro-mobility system now includes multiple private operators of e-scooters and shared bikes, offering thousands of vehicles available across the city. This study focused on e-scooters, as they were introduced later than bike-sharing systems and have received comparatively less research attention. The basic analytical unit for community detection is the CBG. CBGs provide rich socioeconomic and demographic data from the American Community Survey (ACS), making them ideal for identifying communities and conducting further analysis.

\subsection{e-scooter data}

\begin{figure}[ht!]
\centering\includegraphics[width=1.0\linewidth]{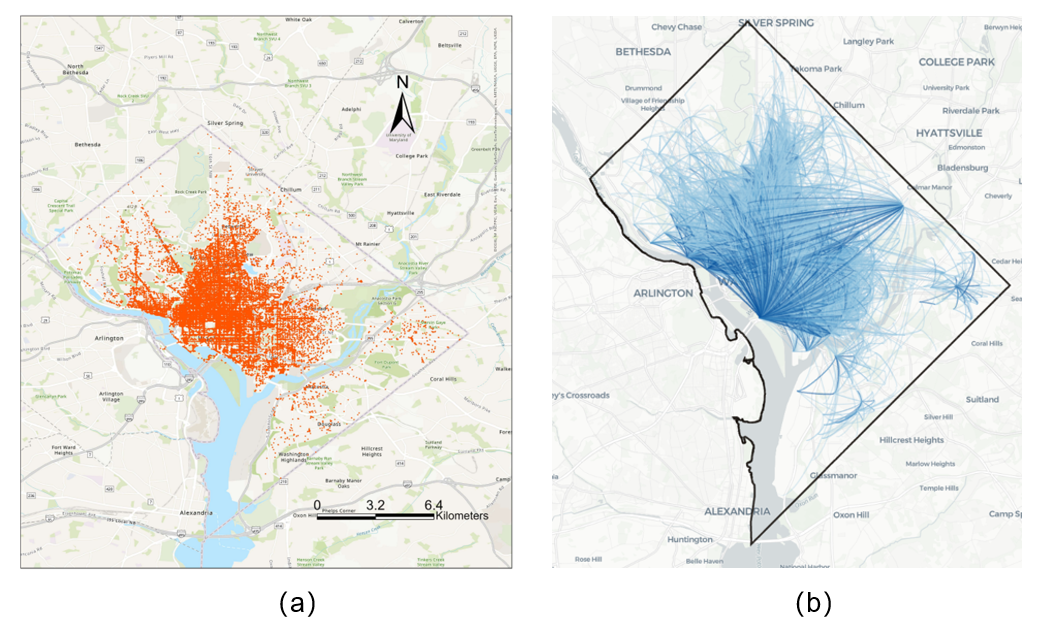}
\caption{The overview of e-scooter data together with OD flows in the study area: (a) the pickup and drop-off locations for e-scooter trips; (b) the OD flows of e-scooter trips (The darker the color, the higher the number of trips). }
\label{fig:01}
\end{figure}

Figure \ref{fig:01} illustrates a snapshot of dockless e-scooter data within our designated study area. By submitting requests to the publicly accessible APIs offered by DDOT at 60-second intervals, we compiled a dataset encompassing nine weeks of dockless e-scooter activity from two vendors (Jump and Bird), spanning from December 13, 2019, to February 14, 2020 \citep{li2022understanding}. We cleaned the data by excluding trips that were likely generated from operator-related redistribution or maintenance activities \citep{mckenzie2019spatiotemporal}. Consequently, a total of 83,002 trips were extracted, each accompanied by timestamps and the coordinates of trip origins and destinations. All trips involved a total of 14,328 e-scooters. Among these trips, 50.60\% had a duration of over 10 minutes, and 11.09\% lasted longer than 30 minutes.

To facilitate community detection, we spatially joined the e-scooter OD to CBGs. As the analysis focused on interconnections among CBGs, self-loop trips (i.e., trips that start and end within the same CBG) were excluded. After applying additional data refinement procedures based on thresholds for speed, duration, and distance, the dataset was finalized to include 62,246 valid shared e-scooter trips. The utilized e-scooter trip data provides near-comprehensive spatial coverage of the study area. Out of all the CBGs, only three lack any recorded e-scooter trips, and these are situated at the periphery of the study area.

\subsection{Ancillary data}

We obtained land use, POIs and census data to understanding the socioeconomic features of the detected communities from e-scooter system.

The land use data for the study area were obtained from the Open Data DC platform \citep{opendatadc2021} and classified it into five land use categories: commercial, residential, industrial, public and recreational, and open space and parks. For detailed information about the classification methods, please refer to \citep{li2022understanding}.

POI data for Washington, D.C. were sourced from Foursquare, a location-based social media platform. Foursquare maintains a comprehensive global database of millions of venues and locations, ranging from restaurants and cafes to parks and landmarks. Developers can utilize Foursquare's API to integrate this data into their applications, especially for identifying nearby POIs within a specified distance. We also collected the locations of bus stops in our study area to analysis the public transit accessibility. A total of 10,044 bus stops witin the D.C. area were extracted from OpenStreetMap (OSM) \citep{openstreetmap, haklay2008openstreetmap}.

Our dataset includes 19,370 venues categorized into nine primary groups: Outdoors \& Recreation, Travel \& Transport, Art \& Entertainment, Food, Residence, Nightlife \& Spot, Shop \& Service, Professional \& Other Places, and Collective \& University. The POI data was further mapped to CBGs to analyze the spatial distribution of POIs in relation to e-scooter usage in each CBG.

We obtained socioeconomic data at the CBG level from \citep{nhgis}, sourced from the 5-year American Community Survey (2016-2020). For our analysis, we focused on several key socioeconomic indicators, including income, employment status, racial demographics.

\section{GIP: Geospatial interaction propagation model}

\subsection{The framework of the GIP}

\begin{figure}[ht!]
\centering\includegraphics[width=1.0\linewidth]{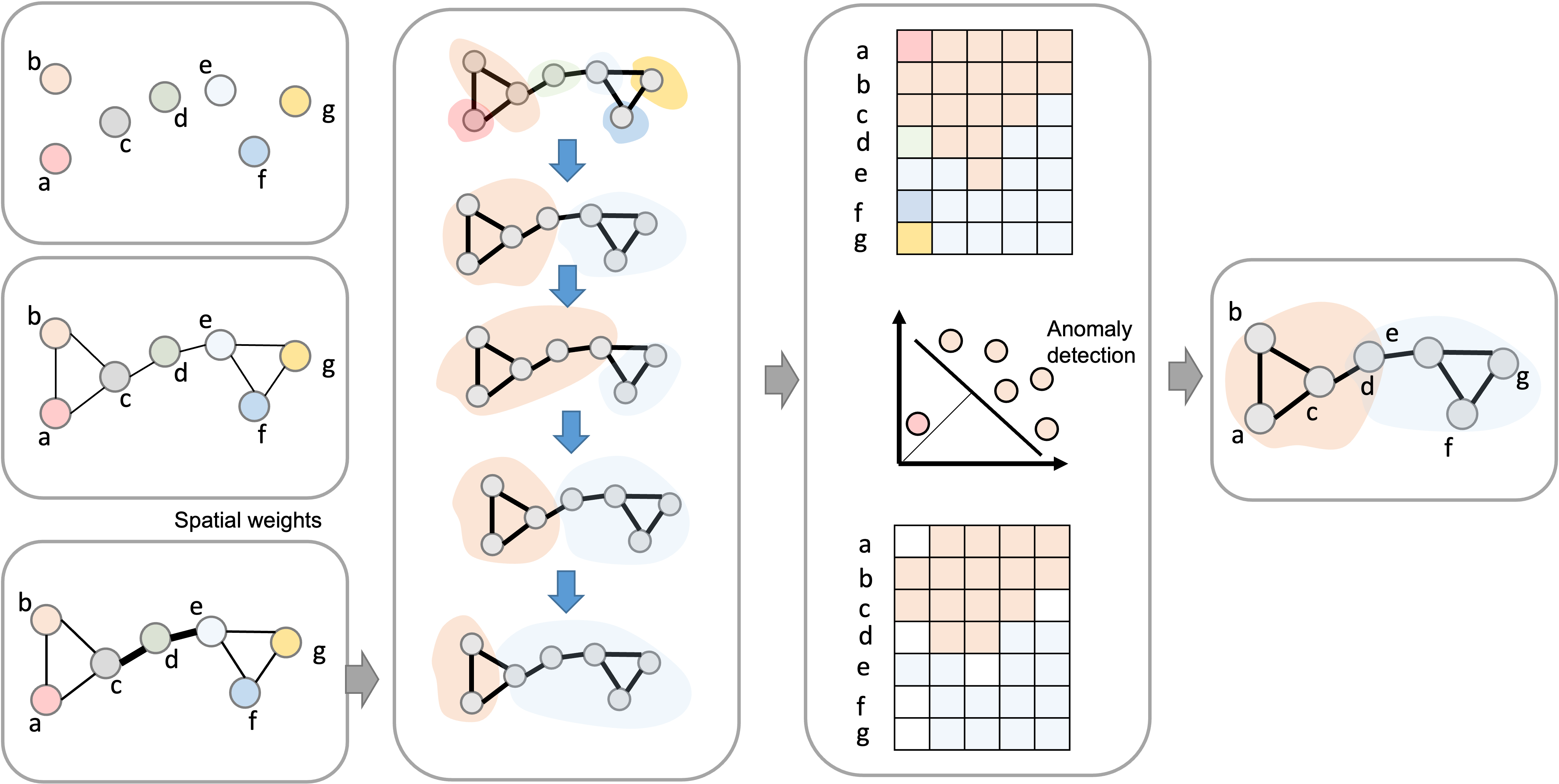}
\caption{The framework of the Geospatial InteractionPropagation (GIP) model.}
\label{fig:02}
\end{figure}

We model human mobility across space as a speaker-listener propagation process. The connections between geographic locations (nodes) are defined by the intensity of human mobility, quantified through the volume of flows and the distance between locations. In this network structure, locations with stronger connections are more likely to form the same spatial community. We hypothesize that human mobility within a community is stronger than between communities, meaning that locations with tighter mobility links will gradually share the same community label. Specifically, the flow of people between locations is treated as the propagation of community labels. A location can either propagate its community label to others (acting as a "speaker") or receive labels from other locations (acting as a "listener"). As the mobility-driven propagation continues, locations with dense connections will gradually converge into the same community. Furthermore, locations that exhibit a significant proportion of multiple different community labels during the propagation process represent the overlapping areas between two communities.

The SLPA method is developed to detect the overlapping communities in social network. However, the SLPA cannot be simply applied in geographical context. First, the SLPA cannot perform community detection on weighted graphs. In geographical context, interactions between different locations varies, which are constrained by their spatial distance and the intensity of human mobility.

Second, during the label propagation process, the SLPA algorithm may affix labels with a low probability of occurrence to the node, leading to the propagation of erroneous labels. So, we need to select a threshold \textit{r} for SLPA, and choosing the \textit{r} value is a challenge. This is because \textit{r} value selection can have a significant impact on community detection results, but there is currently no standardized criterion to guide this choice. Often, the best approach is to experiment with different \textit{r} values and evaluate the effectiveness of community detection. The optimal threshold can be determined by comparing the stability and rationality of the community structure obtained at different \textit{r} values. This process is usually cumbersome and time-consuming. 

To address the above problems, we develop a geospatial interaction propagation model based on the SLAP (Figure \ref{fig:02}). First, we introduced weights when constructing the graph; Second, we improved the SLPA algorithm by changing the selection of the label with the highest count or probability accumulation to the label with the highest weight accumulation for the current node; Thirdly, we propose to automate the threshold calculation for automatic \textit{r}-value selection.

\subsection{Graph construction}

A directed weighted graph can effectively represent a dockless e-scootter system where the graph simulates the correlation of e-scooter flows in distinct census block groups. Additionally, the inclusion of weights on these edges introduces a quantitative dimension, facilitating the portrayal of the intensities or magnitudes characterizing the interplay of e-scooter flows across discrete census block groups.

The graph is denoted as:
\begin{equation}
G = (V, E, W)
\end{equation}
where \(V\) is a set of CBGs, \(E\) is a set of edges, and \(W\) is a matrix of weights. The CBGs, signifying the land cells where e-scooter journeys start or end, are characterized by:
\begin{equation}
V = \{CBG_i\}_{i=1}^{n}
\end{equation}
with \(n\) denoting the total number of CBGs.

The edges in the network are denoted by:
\begin{equation}
E = V \times V = \{e_{ij}\}
\end{equation}
where \(e_{ij}\) is the edge between \(CBG_i\) and \(CBG_j\). As mentioned before, we classified e-scooter trips into CBGs so we can systematically analyze and quantify trips within each CBG.

As Tobler's first law of geography states, "Everything is related to everything else, but near things are more related than distant things" \citep{Tobler1970}. Hence, we define the weights as follows:

\begin{equation}
w_{ij} = \frac{\text{flow}_{ij}}{d_{ij}}
\end{equation}

where \textit{i} and \textit{j} represent the number of CBGs, $w_{ij}$ represents the weight of the edge between $CBG_i$ and $CBG_j$, $flow_{ij}$ is the trip flow between $CBG_i$ and $CBG_j$, and $d_{ij}$ represents the Euclidean distance between the centroid of $CBG_i$ and $CBG_j$. 

After that, we can construct the spatial weight matrix, which is built upon a number of CBGs. The weight matrix of \(M\) CBGs is described by:
\begin{equation}
W = \{w_{ij}\}_{M \times M} = 
\begin{pmatrix}
0 & w_{12} & \dots & w_{1M} \\
w_{21} & 0 & \dots & w_{2M} \\
\vdots & \vdots & \ddots & \vdots \\
w_{M1} & w_{M2} & \dots & 0
\end{pmatrix}
\end{equation}

When we begin by constructing a graph of the data, the nodes are designated as the 571 different CBGs in Washington D.C., and the weights are computed using flow and distance. After constructing the graph, the next step is to apply this structure to a community detection model based on spatial mobility.

\subsection{Geospatial weighted SLPA}

\begin{figure}[ht!]
\centering\includegraphics[width=1.0\linewidth]{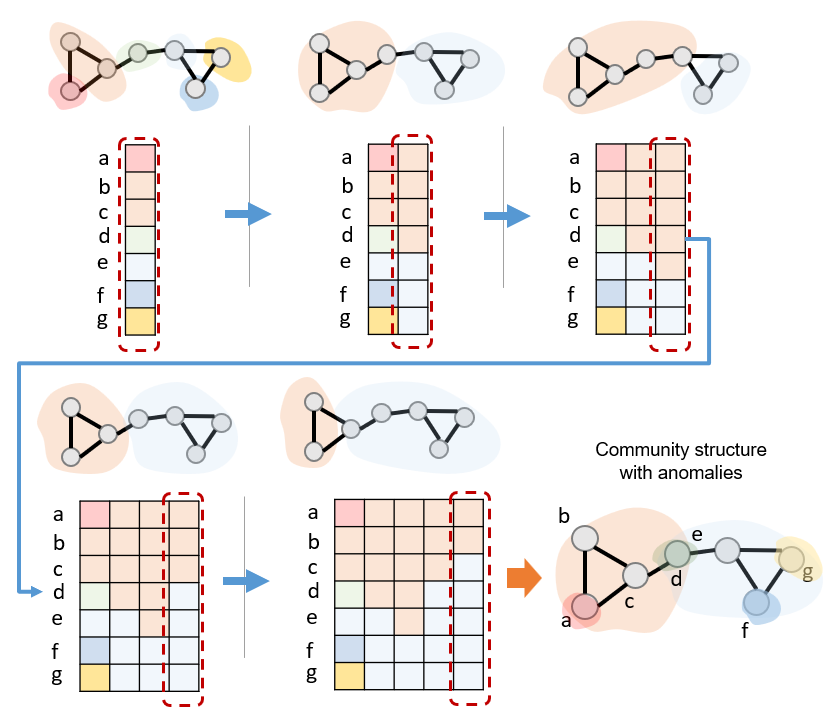}
\caption{The geospatial weighted SLPA}
\label{fig:03}
\end{figure}

Since SLPA cannot perform community detection on weighted graphs and we introduced weights when constructing the graph, we improved the SLPA algorithm by changing the selection of the label with the highest count or probability accumulation to the label with the highest weight accumulation for the current node. The process of the geosaptial weighted SLPA is as shown in Figure \ref{fig:03}. It follows most of the process same as SLPA, but introduce the  

Here we improved the SLPA model with the geographical weight:

\begin{enumerate}
    \item \textbf{Initialization of Node Memory:} Each node is assigned a unique label.
    \item \textbf{Iterative Process (Repeated Until Stop Criterion is Met):}
    \begin{enumerate}
        \item \textbf{Listener Selection:} A node is chosen to act as a listener in each iteration.
        \item \textbf{Label Propagation:} Each neighbor of the selected listener node sends a label based on a specified speaking rule.
        \item \textbf{Label Reception:} The listening node receives a tag from its neighbor’s tag set based on the specified listening rule.
    \end{enumerate}
    \item \textbf{Post-Processing for Community Output:} Finally, post-processing is conducted based on the labels stored in the memories of nodes to generate the output representing the identified communities.
\end{enumerate}

\subsection{Anomaly detection with One-Class SVM}

In the implementation of SLPA for community detection, selecting appropriate threshold pairs is critical. However, determining the optimal threshold typically requires trial and error, as no standardized method exists. To address this, we aim to automate the threshold selection process using anomaly detection techniques. The ASLPAw function from the Python package CDLib \citep{rossetti2019cdlib} employs an isolation forest for anomaly detection within the SLPA model. However, the isolation forest-based approach is computationally intensive and time-consuming.

One-Class SVMs are different from other SVMs in that they have only one class of training samples. The result is "yes" if the sample belongs to the class, and "no" if it doesn't. Therefore, one-class SVMs are often used in outlier detection.

\begin{figure}[ht!]
\centering\includegraphics[width=1.0\linewidth]{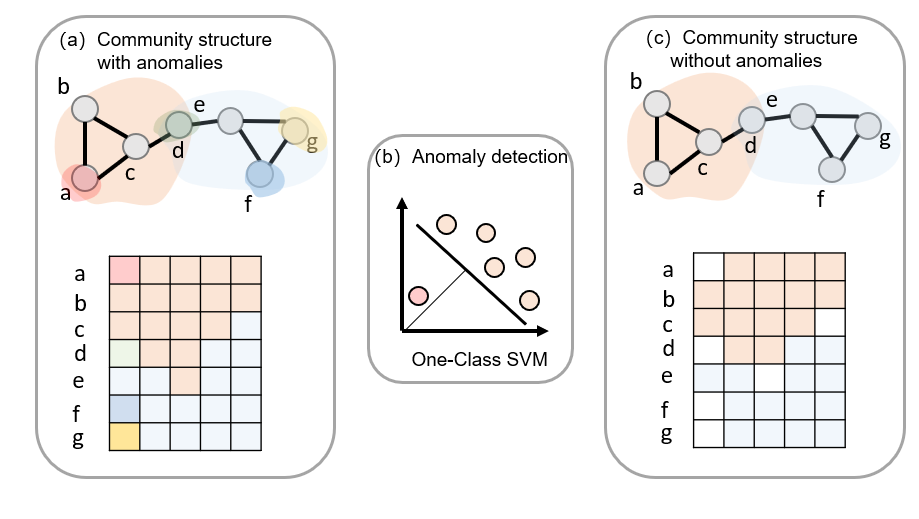}
\caption{Anomaly detection for the community detection results}
\label{fig:04}
\end{figure}

The pesudo-code of GIP is as follows: 

\begin{algorithm}
\caption{Network Evolution Process}
\begin{algorithmic}[1]
\STATE $[n, \text{CBGs}, \text{weights}] = \text{loadNetwork}()$

\STATE \textbf{Stage 1: Initialization}
\FOR{$i = 1$ \TO $n$}
    \STATE $\text{CBGs}(i).\text{Mem} = i$
\ENDFOR

\STATE \textbf{Stage 2: Evolution}
\FOR{$t = 1$ \TO $T$}
    \STATE $\text{CBGs}.\text{ShuffleOrder}()$
    \FOR{$i = 1$ \TO $n$}
        \STATE $\text{Listener} = \text{CBGs}(i)$
        \STATE $\text{Speakers} = \text{CBGs}(i).\text{getNbs}()$
        \STATE $\text{Weight} = \text{weights}(i)$
        \FOR{$j = 1$ \TO $\text{Speakers}.\text{len}$}
            \STATE $\text{LabelList}(j) = \text{Speakers}(j).\text{speakerRule}()$
            \STATE $\text{LabelWeight}(j) = \text{Weight}(j).\text{speakerRule}()$
        \ENDFOR
        \STATE $l = \text{Max\_weighted\_label}(\text{LabelList}, \text{LabelWeight})$
        \STATE $\text{Listener}.\text{Mem.add}(l)$
    \ENDFOR
\ENDFOR

\STATE \textbf{Stage 3: Post-processing using One-Class SVM}
\FOR{$i = 1$ \TO $n$}
    \STATE $\text{Labels} = \text{CBGs}(i).\text{Mem.getLabels}()$
    \STATE $\text{SVM\_Model} = \text{trainOneClassSVM}(\text{Labels})$
    \STATE $\text{CBGs}(i).\text{Mem} = \text{SVM\_Model.filterLabels}(\text{Labels})$
\ENDFOR

\end{algorithmic}
\end{algorithm}

\section{Case study: detecting the overlapping communities of  escooter system with GIP model}

\subsection{Experiment design}

\begin{figure}[ht!]
\centering\includegraphics[width=1.0\linewidth]{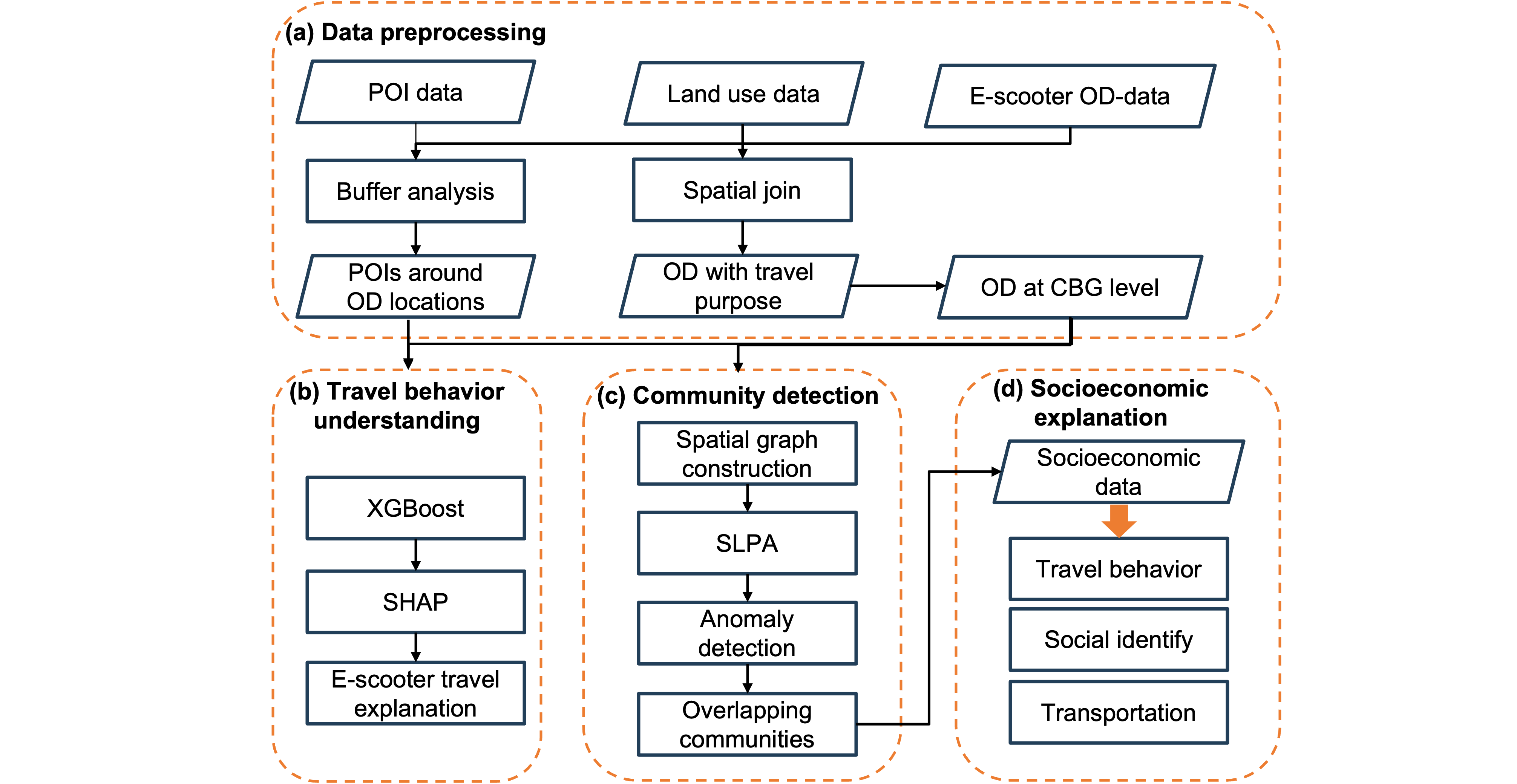}
\caption{The framework of applying GIP for e-scooter community detection}
\label{fig:05}
\end{figure}

We utilized the proposed GIP model to detect the overlapping communities in e-scooter systems at Washington, D.C. The experiment includes four parts (Figure \ref{fig:05}).

First, we conducted data preprocessing. We excluded OD flows with very low frequencies (less than 1) from the OD network. In addition, based on location, we calculated the land use type and POI type for each e-scooter trip’s drop-off point. For POI identification, we referred to the previous study \citep{li2022understanding} and established a 300-meter buffer to capture nearby venues. Next, we matched the e-scooter trips, which contained both land use and POI types, with the CBG of the corresponding pick-up and drop-off points. This allowed us to build a CBG-level OD network for subsequent analysis based on our proposed GIP method.

Second, we applied explainable artificial intelligence (XAI) techniques to interpret the e-scooter trip model \citep{li2022extracting,li2023locally,luo2024understanding}. We built an XGBoost model using explanatory variables (as shown in Appendix Table 1) to predict the number of drop-offs (D). Then, we employed the SHapley Additive exPlanations (SHAP) to explain the predictions made by the XGBoost model.

Third, the GIP model was used to identify overlapping communities. We used time consumption and modularity to verify performance of the GIP.

\begin{equation}
Q = \frac{1}{2m} \cdot \sum_{ij} \left[ A_{ij} - \frac{k_i \cdot k_j}{2m} \right] \delta(C_i, C_j)
\end{equation}

where \textit{Q} is the value of modularity degree. \textit{m} is the sum of the total weights of all edges in the network. \textit{i} and \textit{j} denote the nodes in the network. $A_{ij}$ is the edge weight between node i and node j if the connection exists. $k_i$ is the degree of node \textit{i}, i.e., the sum of weights of the edges that are connected to node \textit{i}. $\delta(C_i, C_j)$ is the community assignment function, which returns 1 if node \textit{i} and node \textit{j} belongs to the same community, and 0 otherwise. $C_i$ and $C_j$ denote the communities to which node \textit{i} and node \textit{j} belong, respectively.

While the ASLPAw model incorporates the SLPA model with Isolation Forest-based anomaly detection, it does not account for geographical weighting. To ensure a fairer comparison of the performance between GIP and existing methods, we introduced geographical weighting into ASLPAw, developing the geospatially weighted ASLPAw. We then compared our proposed GIP model with the geospatially weighted ASLPAw in terms of computational efficiency and modularity.

Finally, based on the identified community detection results, we analyzed the socioeconomic differences between overlapping and non-overlapping communities. Our analysis focused on three key aspects: the land use types of e-scooter trip destinations, the local socioeconomic and demographics (such as income and race), and the local transportation services.

\subsection{Parameter optimization}

As for parameter optimization, the One-Class SVM algorithm in scikit-learn contains three fundamental parameters crucial for configuring its behavior. 'kernel' specifies the type of kernel used in the algorithm. If not given, 'rbf' will be used. The parameter 'nu' is critical as it specifies an upper bound on the fraction of training errors and a lower bound of the fraction of support vectors. The 'gamma' parameter is the Kernel coefficient for 'rbf', 'poly' and 'sigmoid'. In our approach, default values are used for all parameters except the ‘nu’ parameter. 

\begin{figure}[ht!]
\centering\includegraphics[width=1.0\linewidth]{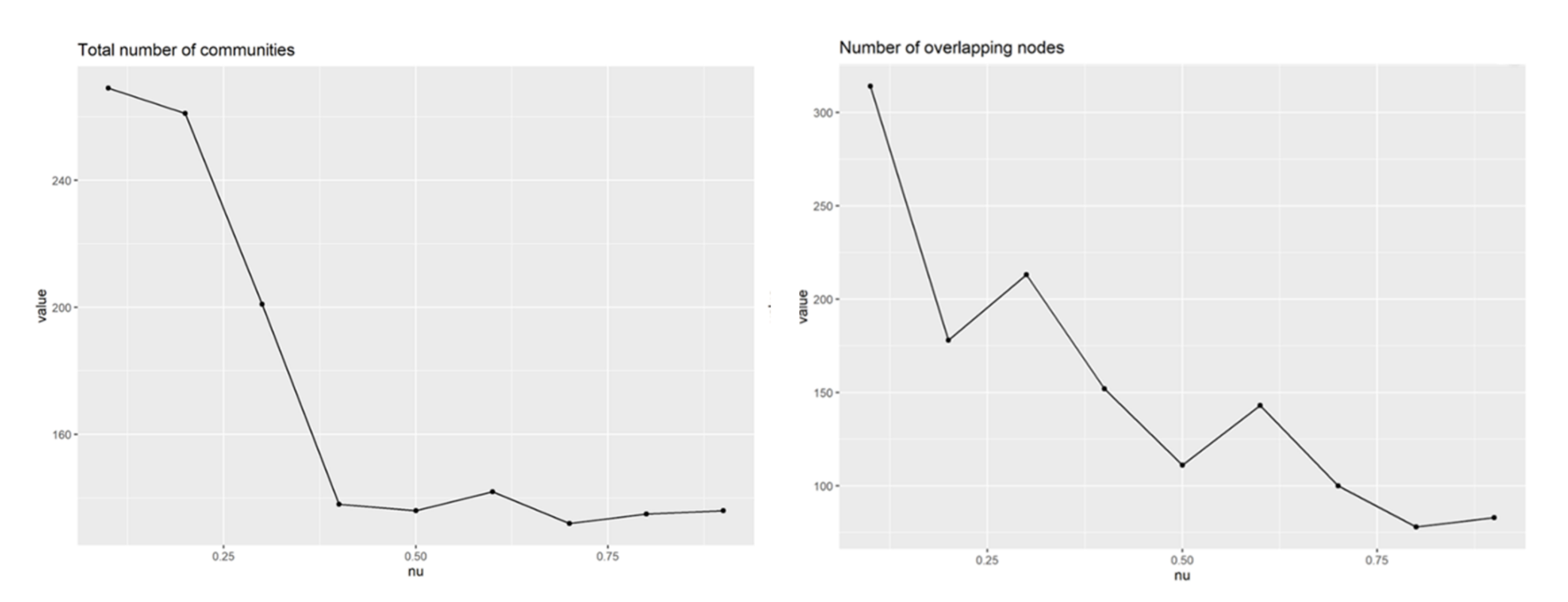}
\caption{Parameter optimization}
\label{fig:12}
\end{figure}

The ‘nu’ parameter must be explicitly defined because the ‘nu’ value has a significant impact on the model. By setting the appropriate ‘nu’ value, we can get a suitable model to remove outliers. The careful selection of ‘nu’ is the key to aligning the One-Class SVM with the desired level of flexibility and robustness in capturing deviations from the norm.  Figure \ref{fig:12} illustrate the results with varying values of ‘nu’ in the range of 0 to 1, with an interval of 0.1. It is observed that the total number of communities tends to stabilize when nu is greater than or equal to 0.4. However, for the number of overlapping nodes, it can be observed from these figures that when ‘nu’ exceeds approximately 0.8, the number of overlapping nodes stabilizes around 80, and the magnitude of change in overlapping nodes also becomes stable. Consequently, we have set the value of nu to 0.8.

\subsection{Understanding the e-scooter travel behavior with explainable AI}

\begin{figure}[ht!]
\centering\includegraphics[width=1.0\linewidth]{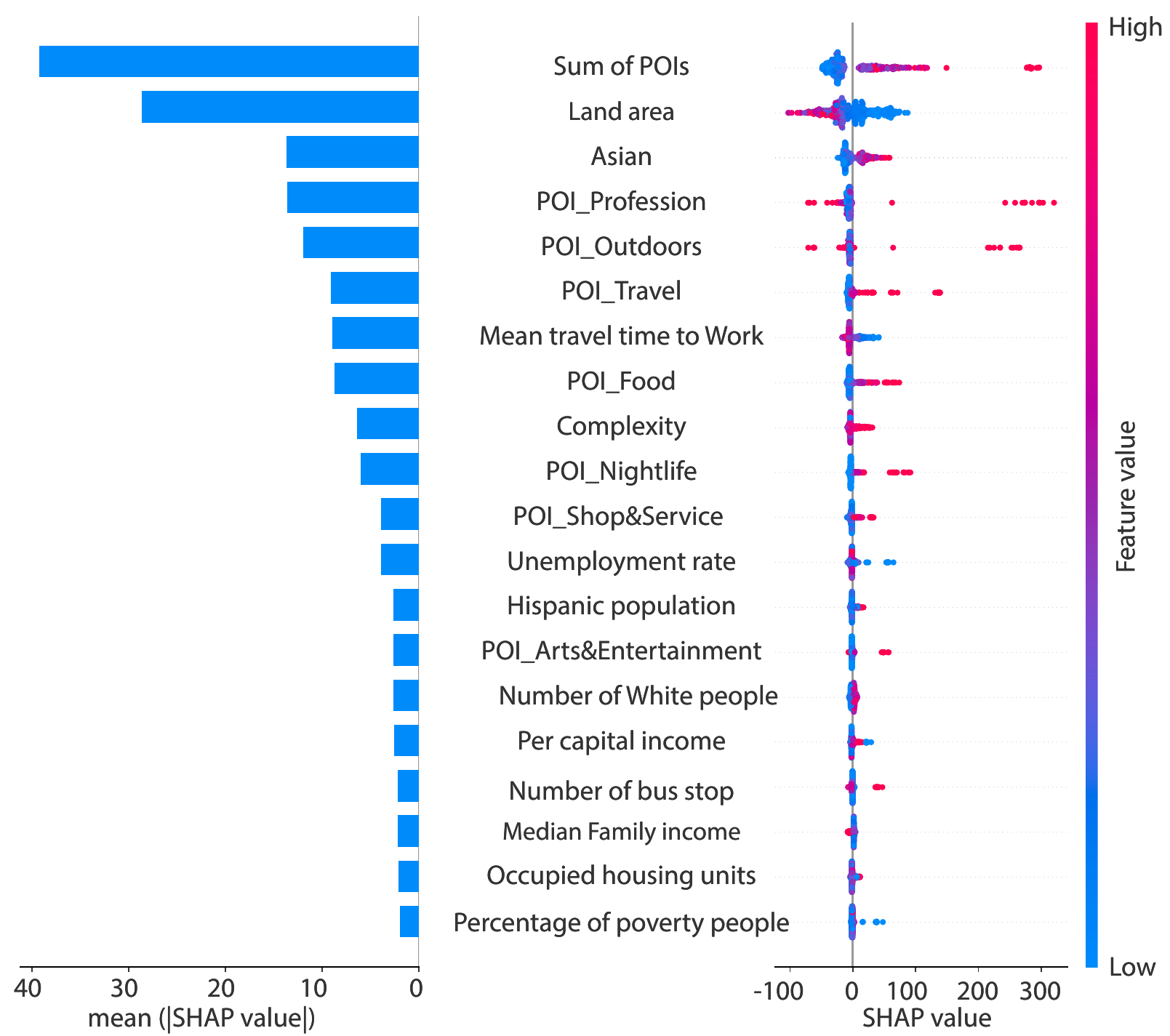}
\caption{The SHAP values of explanatory variables in e-scooter travel behavior analysis, showcasing only the top 20 variables with the highest average SHAP values.}
\label{fig:06}
\end{figure}

Figure \ref{fig:06} presents the results of SHAP explanation of the impact of various factors on e-scooter usage. Our model incorporates variables such as POIs, transportation indicators, demographic data, and socioeconomic indicators. The results show that the sum of POIs and land area are the two most significant factors influencing e-scooter trips. Interestingly, while the sum of POIs positively impacts trip numbers, land area has a negative effect. This indicates that areas with smaller geographic size but a higher density of services attract more e-scooter trips. 

For demographic characteristics, racial composition has a strong impact on e-scooter usage. More White and Asian populations are associated with positive SHAP values, indicating more e-scooter usage in these areas. In contrast, areas with more Black populations shows an opposite pattern. In terms of socioeconomic factors, such as "unemployment rate" and "mean family income," contrasting effects are noted. Lower unemployment rates and higher family incomes correspond to positive SHAP values, suggesting that higher socioeconomic conditions encourage more e-scooter usage.

\begin{figure}[ht!]
\centering\includegraphics[width=1.0\linewidth]{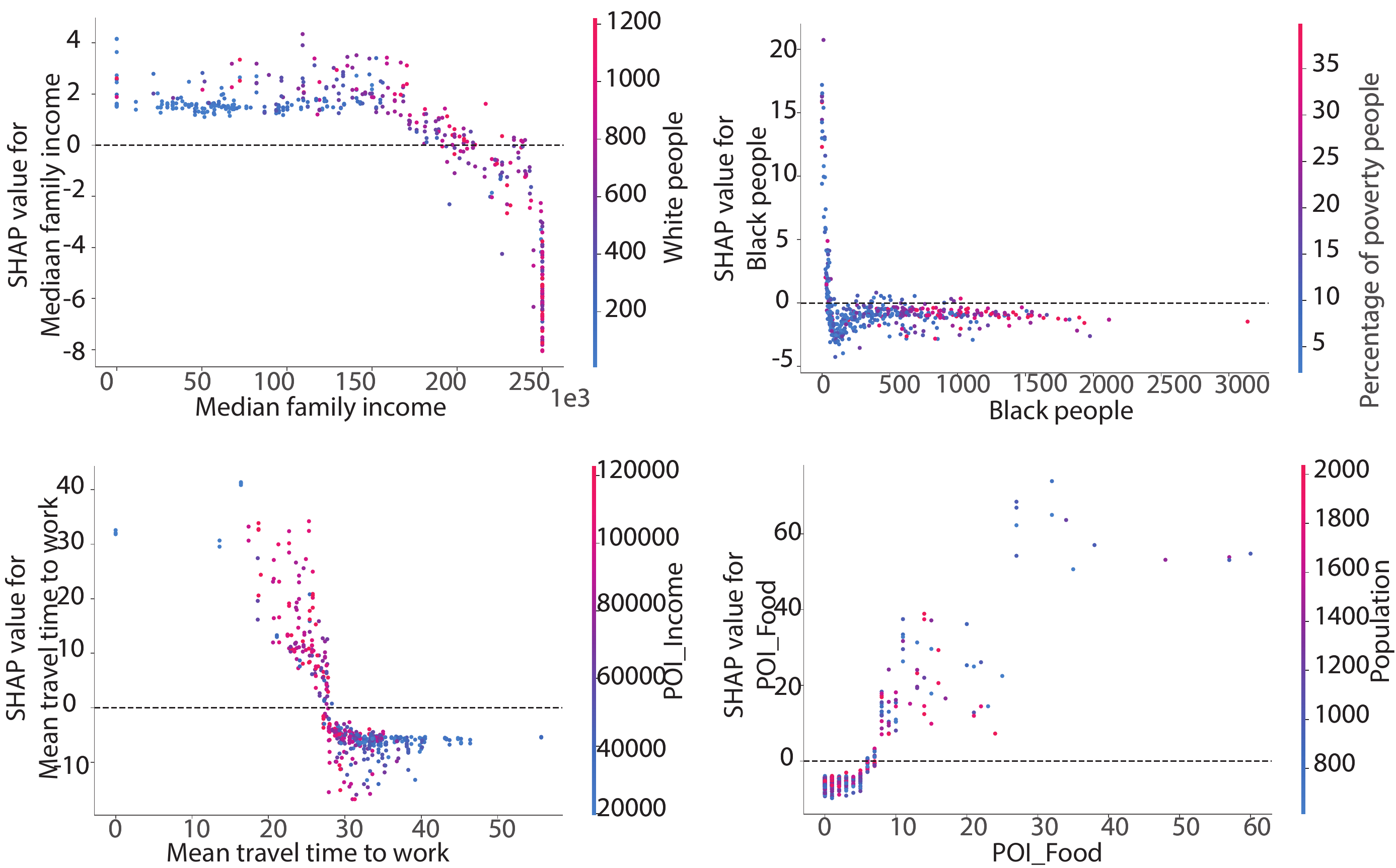}
\caption{The interaction effects of variables to e-scooter travel behaviour: a)Median family income; b) Number of black people; c)Mean travel time to work; d)$POI_food$}
\label{fig:07}
\end{figure}

Figure \ref{fig:07} presents SHAP-based partial dependence plots, highlighting the relationships between specific features and e-scooter usage. Figure \ref{fig:07}a shows a non-linear relationship between median family income and e-scooter usage: usage increases with income initially but plateaus and declines at very high levels, with this effect modulated by the number of white residents. Figure \ref{fig:07}b reveals a sharp decline in e-scooter usage as the number of Black residents rises, modulated by poverty levels.This figure shows that CBGs with a higher number of Black residents tend to have a reducing effect on trips. These areas are also often characterized by a high proportion of poverty.
Figure \ref{fig:07}c indicates a significant negative correlation between mean commute time and e-scooter usage, with patterns varying in higher-income areas. In Figure \ref{fig:07}d, an increase in food-related POIs initially boosts e-scooter usage, but the effect diminishes beyond a certain point, with total population influencing earlier saturation in denser areas.

\subsection{Overlapping community detection results}

\subsubsection{Model evaluation}

The comparison between geospatially weighted ASLPAw and GIP model, as shown in the Table \ref{table2}, highlights significant differences in both performance and efficiency. Our model consistently achieves higher modularity values across the 10 runs, indicating better community detection performance. Additionally, our model significantly outperforms ASLPAw in terms of computational efficiency, with running times approximately an order of magnitude lower. This demonstrates that our model not only provides more accurate community structures but also does so with much greater speed.

\begin{table}[h]
\centering
\caption{Performance Comparison between geospatially weighted ASLPAw and GIP Model}
\begin{tabular}{|c|c|c|c|c|}
\hline
 & \multicolumn{2}{c|}{geospatially weighted ASLPAw} & \multicolumn{2}{c|}{GIP Model} \\ \hline
 & Time (s) & Modularity & Time (s) & Modularity \\ \hline
Average & 106.5392 & 0.298227 & 7.92115 & 0.344703 \\ \hline
SD & 3.137367 & 0.013965 & 0.160305 & 0.069829 \\ \hline
\end{tabular}
\label{table2}
\end{table}

\subsubsection{Distribution of overlapping communities}

\begin{figure}[ht!]
\centering\includegraphics[width=1.0\linewidth]{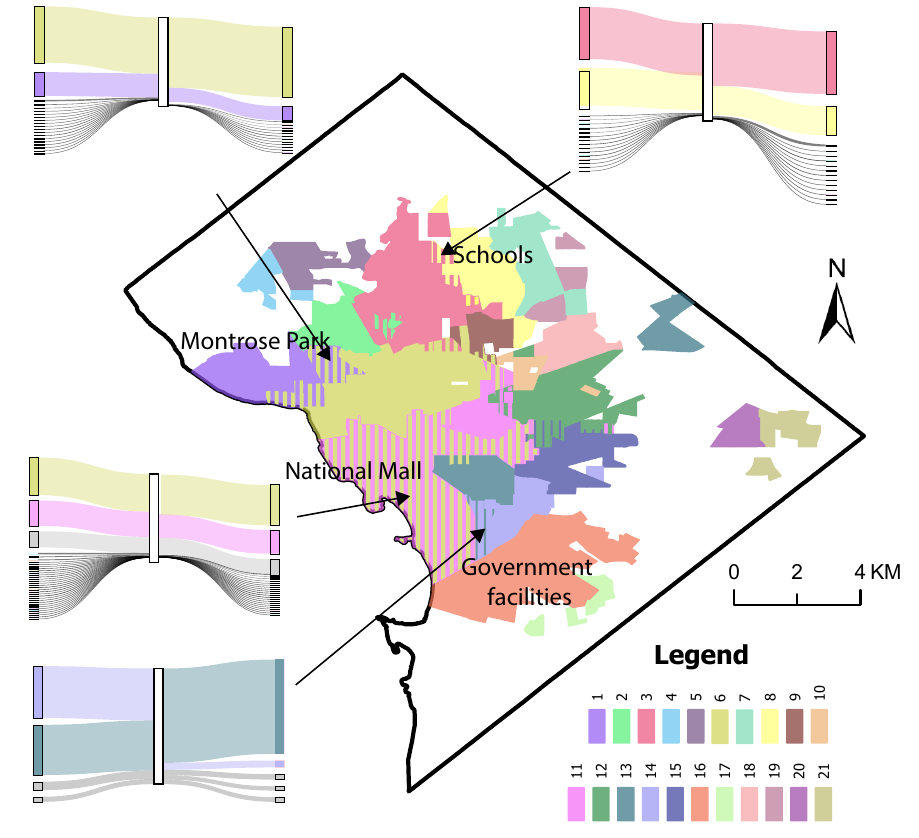}
\caption{Overlapping communities of e-scooter sharing system in Washington D.C.: CBGs in different colors represent their respective communities. Overlapping CBGs across multiple communities are visualized with diagonal hatching. The sankey diagram to the overlapped CBGs shows the starting points of e-scooter trips visiting these CBGs and the endpoints of trips departing from them.}
\label{fig:08}
\end{figure}

Figure \ref{fig:08} illustrates the community distribution of e-scooter usage as revealed by the model. We identified a total of 86 communities, with 21 communities having more than 4 CBGs. Additionally, we discovered that 41 CBGs belong to more than one community, indicating overlapping locations. For these overlapping locations, we found that they often include parks, squares, and various schools. For example, the CBG where the National Mall is located overlaps between two adjacent communities (11 and 6). This area not only contains political buildings such as the White House and the Department of Defense but also landmarks like the Lincoln Memorial and the Washington Monument. Its dual significance as a political and tourist hub attracts people from diverse origins, contributing to it becoming an overlapping area. Furthermore, in Figure \ref{fig:08}, we selected several key overlapping areas to display the proportion of e-scooter trips originating from different communities, which help us better understand the motivations behind e-scooter usage in different areas and can guide more effective planning of e-scooter networks.

The Table \ref{table1} provides statistics for both overlapping and non-overlapping areas. As can be seen from the table, there are 41 overlapping CBGs and 530 non-overlapping CBGs. Compared to the non-overlapping areas, the overlapping areas have a higher average land and water area, a greater number of trips, and a slightly smaller population. This suggests that overlapping areas also have more travel activity and serve as key connective hubs within the city.

\begin{table}[h]
\centering
\caption{Comparison of characteristics between overlapping and non-overlapping CBGs}
\begin{tabular}{|l|c|c|}
\hline
 & Overlap & Non-Overlap \\ \hline
CBG Count & 41 & 530 \\ \hline
Average Land Area (m$^2$) & 325,781.6 & 273,507.7 \\ \hline
Average Water Area (m$^2$) & 143,068.2 & 24,233.9 \\ \hline
Average Population (persons) & 1,046.4 & 1,220.1 \\ \hline
Average Trips (trips) & 265.6 & 96.9 \\ \hline
\end{tabular}
\label{table1}
\end{table}

\subsection{Socioeconomic features explained by overlapping community}

\subsubsection{Travel behaviour}

\begin{figure}[ht!]
\centering\includegraphics[width=1.0\linewidth]{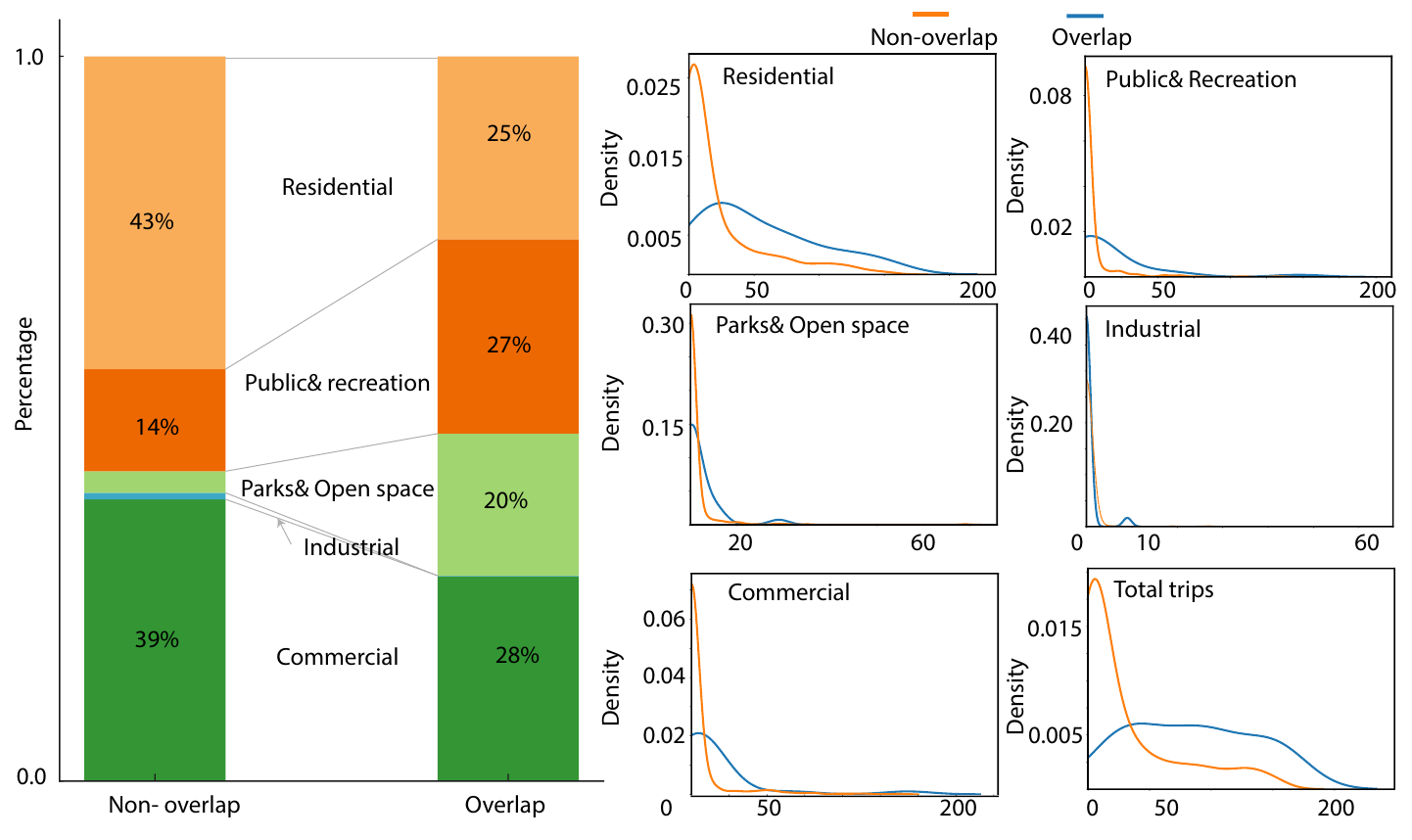}
\caption{Overlapping communities of e-scooter sharing system in Washington D.C.}
\label{fig:09}
\end{figure}

Based on the previous data preprocessing, each trip is associated with the land use type of its destination. Therefore, we analyzed the land use distribution of all trips in both overlap and non-overlap areas. Figure \ref{fig:09} shows that in non-overlap areas, 43\% of e-scooter trips are destined for residential areas, while this figure is only 25\% in overlap areas. Additionally, we found that a significantly higher proportion of e-scooter trips in overlap areas are directed towards public and recreation areas (27\%) and parks and open spaces (20\%). The Kernel density estimation (KDE) plots illustrate the trip density distribution across different land use categories at the CBG level, indicating that high-density trips in commercial, public, and recreational categories are more common in overlap areas. Furthermore, the KDE results for total trips suggest that CBGs with higher e-scooter trip volumes are more likely to be located in overlap areas.

Table \ref{table3} presents the statistic results of trips for different land use. It reveals that overlap areas generally exhibit higher mean and maximum values for trips compared to non-overlap areas, indicating greater activity and trip generation within these regions. This suggests that overlapping areas are significant hubs for various trip purposes due to the higher density of points of interest.

\begin{table}[h]
\centering
\caption{Comparison of Characteristics for Overlap and Non-Overlap Categories}
\begin{tabular}{|l|l|c|c|c|c|}
\hline
 &  & Mean & Max & Min & SD \\ \hline
\multirow{2}{*}{Commercial} & Overlap & 75.05 & 932 & 0 & 176.39 \\ \cline{2-6} 
 & Non-Overlap & 37.67 & 2710 & 0 & 170.91 \\ \hline
\multirow{2}{*}{Industrial} & Overlap & 0.22 & 9 & 0 & 1.41 \\ \cline{2-6} 
 & Non-Overlap & 0.86 & 158 & 0 & 8.12 \\ \hline
\multirow{2}{*}{Parks and Open Space} & Overlap & 52.05 & 2044 & 0 & 318.92 \\ \cline{2-6} 
 & Non-Overlap & 2.89 & 328 & 0 & 19.2 \\ \hline
\multirow{2}{*}{Public and Recreation} & Overlap & 71.22 & 1696 & 0 & 266.07 \\ \cline{2-6} 
 & Non-Overlap & 13.64 & 529 & 0 & 55.99 \\ \hline
\multirow{2}{*}{Residential} & Overlap & 67.05 & 228 & 0 & 59.42 \\ \cline{2-6} 
 & Non-Overlap & 41.83 & 506 & 0 & 74.58 \\ \hline
\multirow{2}{*}{Total Trips} & Overlap & 265.59 & 3816 & 18 & 597.6 \\ \cline{2-6} 
 & Non-Overlap & 96.9 & 2824 & 0 & 209.82 \\ \hline
\end{tabular}
\label{table3}
\end{table}

\subsubsection{Social identify}

\begin{figure}[ht!]
\centering\includegraphics[width=1.0\linewidth]{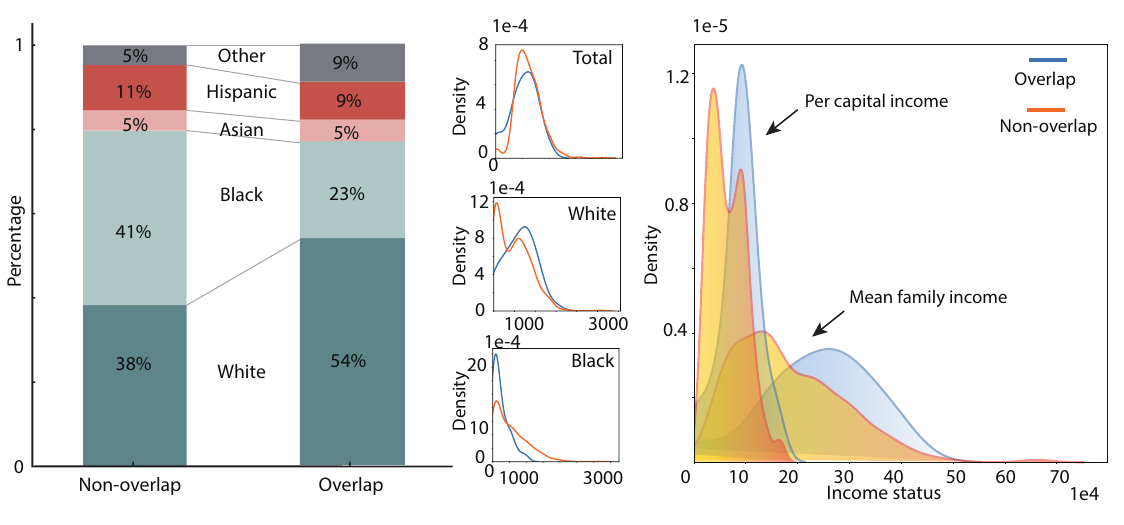}
\caption{Overlapping communities of e-scooter sharing system in Washington D.C.}
\label{fig:10}
\end{figure}

Figure \ref{fig:10} shows the racial composition in overlap and non-overlap areas. While there is no significant difference in total population (Figure \ref{fig:10}b), overlap areas have a significantly higher proportion of White residents (54\%) and a notably lower proportion of Black residents (23\%) compared to non-overlap areas. We also analyzed income levels within the overlap and non-overlap areas. Figure \ref{fig:10}e indicates that residents in overlap areas have significantly higher per capita income and mean family income. This suggests that areas overlapping multiple communities are more likely to be predominantly white people and rich residents.

The relationship between racial composition and community overlap may be influenced by the correlation between race and income. To verify this and further explore the independent effect of racial composition on community overlap, we applied the residualization method \citep{forrest1983residualization}. Specifically, we regressed the White Ratio, Black Ratio, and Asian Ratio on Per Capita Income to calculate the residuals, thereby removing the influence of income on racial composition. These residuals were then included in a logistic regression model to assess their direct impact on community overlap.

The logistic regression model yielded a Pseudo R² of 0.0253, indicating weak overall explanatory power. This suggests that the relationship between racial composition and community overlap is primarily driven by its correlation with income. The regression coefficient for the White Ratio is 0.1820, indicating that an increase in the White population may slightly increase the likelihood of community resource overlap, though this effect is not statistically significant (p = 0.880). The Black Ratio shows a negative effect with a regression coefficient of -1.1890, but this result is also not significant (p = 0.196). In contrast, the Asian Ratio has a regression coefficient of -12.7983, which is statistically significant (p = 0.048), suggesting that, after controlling for income, a higher Asian population ratio significantly decreases the likelihood of community resource overlap.

\subsubsection{Transportation}

\begin{figure}[ht!]
\centering\includegraphics[width=1.0\linewidth]{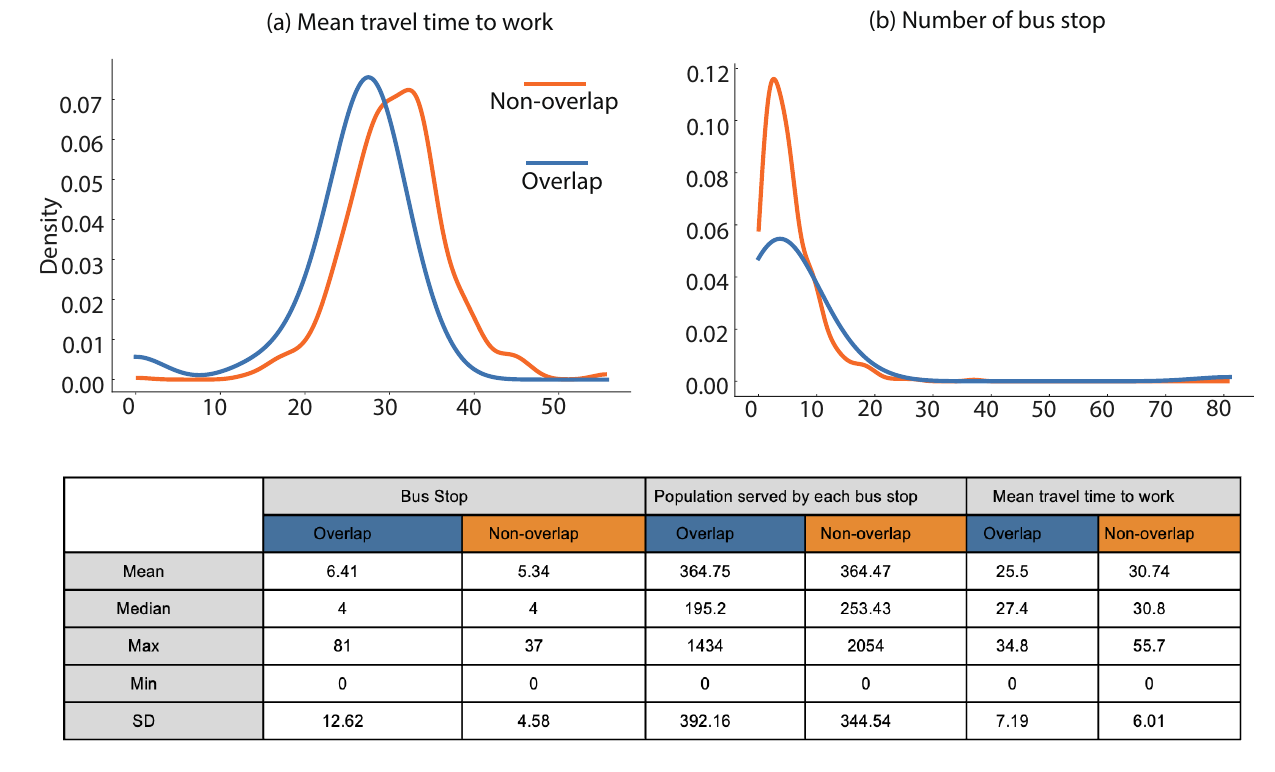}
\caption{Overlapping communities of e-scooter sharing system in Washington D.C.}
\label{fig:11}
\end{figure}

We analyzed the differences in transportation services between overlap and non-overlap areas, represented by mean commute time and the number of bus stops. Figure \ref{fig:11}a shows that commute times in overlap areas (blue line) are generally shorter, with a mean of around 25.5 minutes, while the mean commute time in non-overlap areas (red line) is approximately 30.74 minutes. This suggests that residents in overlap areas are either closer to their workplaces or benefit from better public transit services, resulting in shorter commute times. Figure \ref{fig:11}b shows that the average number of bus stops in overlap areas is 6.41, compared to 5.34 in non-overlap areas. A surprising result (Figure \ref{fig:11}c) is that although the average population served per bus stop is almost identical between overlap and non-overlap areas, at 364.75 and 364.47 people, respectively, the median and maximum number of people served per bus stop in non-overlap areas are significantly higher than in overlap areas, indicating that public transportation resources are more strained in non-overlap areas.

\section{Discussion}

\subsection{The label propagation process in mobility network}

In our study, we propose a new method for detecting overlapping communities. By introducing geographical weights into the SLPA model and combining it with one-class SVM, we developed the GIP model. We applied this model to e-scooter usage data from Washington, D.C., and the results show that it outperforms previous methods in defining community boundaries and identifying overlapping communities.

In the GIP model, human mobility is modeled as a Speaker-Listener propagation process, which represents a high-level abstraction of the interactions between people and places. By treating human mobility as the propagation of community labels, the model reveals the connections between locations and how people play a key role in these connections. This propagation process demonstrates several layers of social interaction mechanisms:

First, human mobility depends on the match between people’s preferences and the functional attributes of destinations. Each location has unique socio-economic functions, such as business areas, residential zones, or cultural sites, which influence people’s choices and mobility patterns. These locations attract similar groups of people and, through daily movements and interactions, shape the movement patterns. Through the Speaker-Listener mechanism, community labels reflect people’s behavior patterns and mobility choices in geographic space, showing how people use and depend on different locations' functions. As mobility continues, frequently visited places gradually form communities, revealing how social groups interact around geographic nodes.

Second, the model highlights the mutual shaping between human flow and locations. Human mobility patterns are influenced by location characteristics, but they also reshape the functionality and importance of places. For example, a commercial area attracting heavy traffic might expand its boundaries, drawing in more resources and facilities. In turn, locations with frequent human movement develop into more functional and diverse communities. The dynamic roles of “Speaker” and “Listener” in the model show that places are both receivers and transmitters of information, with their interactions constantly adjusting through the ongoing flow of information and community labels.

Finally, this process of spatial interaction also explains the formation of social isolation. Through the bidirectional interaction between people and places, locations with similar labels and social status establish stronger connections in space. Locations with more resources and socio-economic functions, like business hubs or transportation centers, are more likely to become the overlap between multiple communities, acting as key connection points. On the other hand, peripheral areas with fewer resources and less infrastructure often experience lower mobility, becoming isolated communities with limited external connections. This process of community label aggregation through mobility reveals how social inequality in urban space is amplified and reinforced by the interactions between human flow and locations.

\subsection{Overlapping communities in the e-scooter system}

We applied the GIP model to Washington D.C.'s e-scooter system to identify the distribution of overlapping communities and the underlying socio-spatial dynamics. The results show that overlapping communities play a key role in meeting public recreation needs, particularly the use of public spaces like parks. These areas provide convenient transportation options and a variety of activities, especially appealing to white groups. The percentage of white residents in overlapping communities is 54\%, compared to only 38\% in non-overlapping communities. This difference highlights that the convenience of e-scooter usage is mainly enjoyed by white residents, revealing an unequal adoption of this new transportation system across different racial groups. In addition to differences in demographics, the average income in overlapping communities is significantly higher than in non-overlapping areas, further emphasizing the appeal of e-scooters to wealthier groups. Affluent individuals have greater access to diverse travel options like e-scooters, which is not as common in economically disadvantaged non-overlapping communities. Moreover, overlapping communities benefit from more developed transportation services, with higher bus stop density and shorter commute times. This means residents in these areas have more flexible travel options. In contrast, residents in non-overlapping areas experience longer commutes and fewer public transit resources, further increasing transportation inequality in these regions.

It is important to note that e-scooter trips may be influenced by factors unrelated to urban structure, such as pricing policies \citep{kang2024user}. For instance, additional fees may be charged after a certain duration (e.g., 30 minutes), which can affect user behavior. Additionally, our data was collected during the winter season, which could potentially reduce both the duration and distance of trips \citep{shah2023people}. We will explore the impact of these factors in future research, including developing dynamic community detection methods that account for the temporal patterns of e-scooter usage.

\section{Conclusion}

This study proposes a method for detecting overlapping communities based on e-scooter mobility networks. By introducing geographical weights into the SLPA model and integrating it with a one-class SVM, we developed the GIP model. We applied the model to e-scooter data from Washington, D.C., and the results demonstrate that the GIP model outperforms previous methods in both accuracy and computational efficiency. The GIP model conceptualizes human mobility as a "Speaker-Listener" propagation process, uncovering the complex interactions between people and places and illustrating how mobility shapes community structures. Our findings reveal that overlapping communities tend to emerge in areas with high public recreation demand and superior transportation infrastructure. These areas are often inhabited by wealthier who enjoy greater mobility options and better access to resources. In contrast, non-overlapping communities are typically characterized by lower-income residents, longer commutes, and limited public transportation services. This study provides insights for urban planners and policymakers, helping them balance the needs of diverse social groups and prevent the exacerbation of social segregation through the introduction of new transportation technologies. Future research could extend the GIP model to other human mobility systems to assess its generalizability.

\section{Disclosure Statement}

No conflict of interest exists in this manuscript, and the manuscript was approved by all authors for publication.

\baselineskip12pt
\bibliographystyle{elsarticle-harv} 
\bibliography{02_references.bib}

\section*{Appendix}

\begin{table}[htb]
\centering
\caption{Variables and descriptions of features used in the model.}
\small 
\begin{tabular}{|p{5cm}|p{10cm}|} 
\hline
\textbf{Variable} & \textbf{Description} \\ \hline
\multicolumn{2}{|l|}{\textbf{POI Attributes}} \\ \hline
Number of POIs & Number of POIs: arts \& entertainment, college, food, nightlife, outdoors, profession, residence, shop \& service, travel \\ \hline
Sum of POI & Total number of POIs \\ \hline
Complexity & Complexity index based on POI \\ \hline
\multicolumn{2}{|l|}{\textbf{Racial Attributes}} \\ \hline
Proportion of Race Groups & Proportion of White, Black, Asian, Hispanic populations \\ \hline
White/Black & Ratio of White to Black population \\ \hline
\multicolumn{2}{|l|}{\textbf{Socioeconomics}} \\ \hline
Total Housing Units & Total housing units \\ \hline
Occupied Housing Units & Occupied housing units \\ \hline
Vacant Housing Units & Vacant housing units \\ \hline
Employment Status & Employment (16+ years old) \\ \hline
Unemployment Rate & Unemployment rate (\%) \\ \hline
Median Family Income & Median family income (USD) \\ \hline
Mean Family Income & Mean family income (USD) \\ \hline
Per Capita Income & Per capita income (USD) \\ \hline
Below Poverty Line & \% population below poverty line \\ \hline
\multicolumn{2}{|l|}{\textbf{Transportation and land use}} \\ \hline
Bus Stop & Number of bus stops \\ \hline
Population per Bus Stop & Population served per bus stop \\ \hline
ALAND & Land area (sq km) \\ \hline
AWATER & Water area (sq km) \\ \hline
Mean Travel Time to Work & Average travel time (minutes) \\ \hline
\end{tabular}
\normalsize %
\end{table}

The complexity index of POIs presented was computed for each CBG using the following equation:

\begin{equation}
    H(X) = - \sum p(x) \cdot \log p(x)
\end{equation}

where \( p(x) \) represents the proportion of POI type x within the given CBG.


\end{document}